\documentclass[a4paper,UKenglish, cleveref, autoref]{lipics-v2021}
\usepackage{needspace}

\hideLIPIcs
\nolinenumbers

\newcommand\condd{\textrm{CondDeg}_\alpha}
\newcommand\cond{\textrm{Cond}_q}

\title{Condense to Conduct and Conduct to Condense}

\author{Tomasz Kazana} {University of Warsaw, Poland}{tkazana@mimuw.edu.pl}{https://orcid.org/0000-0002-3191-3262}{}
	
\authorrunning{T. Kazana}

\Copyright{Tomasz Kazana}



\begin{document}

\maketitle

\begin{abstract}
In this paper, we present the first explicit examples of low-conductance permutations. The notion of conductance of permutations was introduced by Dodis et al. in ``Indifferentiability of Confusion-Diffusion Networks'', where the search for low-conductance permutations was first initiated and motivated. As part of our contribution, we not only provide these examples, but also offer a general characterization of the problem: we show that low-conductance permutations are equivalent to permutations possessing the information-theoretic properties of Multi-Source-Somewhere-Condensers, a specific variant of somewhere condensers.
\end{abstract}


\section{Introduction}

\subsection{Background}

The starting point of this paper is the concept of \textit{permutation conductance}, introduced by
Yevgeniy Dodis, Martijn Stam, John Steinberger, and Tianren Liu in their seminal paper
\textit{Indifferentiability of confusion-diffusion networks}
\cite{dod}.

Their work focuses on the theoretical analysis of the indifferentiability security of confusion-diffusion networks, investigating how many rounds of certain block cipher and hash function constructions suffice to ensure adequate security. A central primitive in their analysis is a particular permutation, for which the parameter they introduced, known as conductance, is expected to be small. The results of this paper (e.g. Theorem \ref{thm-main}) have a direct impact on these applications, as outlined in Section \ref{imp-crypto}. For a detailed discussion of the motivation and the direct connections between permutations with suitable conductance and the security of confusion-diffusion based ciphers, we refer to \cite{dod}. In this work, however, the conductance of a permutation is primarily treated as a combinatorial concept, situated within the broader framework of the combinatorics of data structures.

From a formal perspective, a permutation $\pi: \{ 0,1\}^{nw} \to \{ 0,1\}^{nw}$ is a bijection on $nw$-bit strings. We view each input and output as $w$ consecutive blocks of $n$ bits each; we call $w$ the \emph{dimension} and $n$ the \emph{block length}. This partition only specifies how we group the bits: $\pi$ can change block contents and mix information across block boundaries. Each such permutation has a certain \textit{degree of conductance}, depending additionally on a parameter $\alpha \in (0, 1)$. This degree (denoted $\condd(\pi)$) is always a real number between $1$ and $w$ and serves as a measure of how random $\pi$ is (the more random the permutation, the closer $\condd$ is to $1$; in particular, an identity permutation has $\condd$ equal to $w$). It turns out that permutations with low $\condd$ (``more random'') are useful in the applications described in \cite{dod}. Fortunately, the authors of \cite{dod} showed (Theorem 3, App. B, probabilistic method) that almost every random permutation has (for parameters $\alpha < \frac{1}{2}-\frac{1}{nw}$, which is sufficient for their applications) $\condd$ very close to $1$ (more precisely: it is no larger than $1 + \frac{\log(3nw)}{\alpha n}$). On the other hand, no explicit constructions have been shown demonstrating how such permutations could be constructed in practice.

In this paper we provide, for the first time, explicit examples of permutations with conductance degree less than $w$ for every dimension $w\geq3$ and every sufficiently large block length $n$, for each $0<\alpha\leq0.8$ (see Theorem \ref{thm-main}). Our constructions are based on what we call Multi-Source-Somewhere-Condenser permutations (see Definition \ref{dfn-sc}), a variant of classical somewhere condensers, such as those defined for example in Definition 1.11 of \cite{raz}.

Moreover, we establish a rather surprising converse result: every permutation with sufficiently low conductance degree must satisfy the Multi-Source-Somewhere-Condenser assumptions. This gives a characterization over a very broad range of parameters; the remaining questions are discussed in Section~\ref{sec-open}.

\subsubsection{Relations to the State-of-the-Art and Previous Results}

In the literature on information theory and its applications in computer science, numerous constructions have been developed to simulate or enhance randomness. These include extractors \cite{raz2005extractors, vadhan2012pseudorandomness}, condensers \cite{ta-Shma2017condenser, doron2016nearly}, expanders \cite{goudarzi2020expander, mihail2017list}, list-decodable and list-recoverable codes \cite{mihail2017list, stinson2006cryptography, doron2016nearly}, samplers \cite{goldreich1993samplers}, and pairwise independent hash functions \cite{bellare1996foundations, shaltiel2011pseudorandomness}. Many of these primitives are closely related or, in a certain sense, formally equivalent (see  \cite{impagliazzo1999hardness, dod, shaltiel2011recent, impagliazzo1989recycle, dodis2009nonmalleable}), and they play a key role in managing randomness in computation.
Within this framework, these constructions have numerous applications in cryptography, including universal hashing, key derivation functions, and randomness extractors for secure computation.

In this work, we show that the notion of a low-conductance permutation, as introduced in \cite{dod}, can be viewed as essentially equivalent to condensers, and thus as another member of this family. Therefore, while combinatorially interesting on its own, it turns out to be equivalent to well-studied primitives. This equivalence is valuable as it enables transferring insights and parameter bounds from classical constructions to low-conductance permutations. At the same time, determining explicit parameters and limitations remains challenging, as problems in this family are inherently difficult. Overall, this perspective establishes a conceptual bridge between the combinatorial properties of permutations and information-theoretic tools, highlighting potential avenues for designing new cryptographic primitives with provable security guarantees.

\subsubsection{Concrete Implications for Cryptography}\label{imp-crypto}

A natural question is whether our results have direct implications for block cipher design. The detailed analysis of the connections between low-conductance permutations and block cipher security is thoroughly discussed in \cite{dod}. Here, we highlight the two main insights.

First, the bounds we obtain are non-trivial (see Theorem \ref{thm-main}), though still far from the ideal case of random permutations (in the probabilistic sense), for which conductance is close to $1$. In contrast, the analysis in \cite{dod} relies on access to such nearly optimal bounds. As a result, formal guarantees can only be established for relatively stringent conductance thresholds. Unfortunately, due to the equivalence with condensers, constructing explicit examples that achieve such bounds remains a major challenge.

Second, it turns out that simple linear permutations (as originally hypothesized in \cite{dod}) are unfortunately insufficient for achieving low conductance. These functions inherently fall short of the required conductance levels because they lack the necessary condenser properties, emphasizing the need for more sophisticated constructions to approach the theoretical optimum.

Overall, while immediate implications for provably secure cipher design are limited, this work provides a rigorous combinatorial and information-theoretic foundation, clarifying the limits and potential strategies for constructing robust block ciphers.

\subsubsection{Final Remarks}

To conclude the introduction, we highlight how this problem exemplifies the non-obvious interconnections between several fields in theoretical computer science and mathematics. The problem originally arises in the theoretical analysis of symmetric ciphers -- a relatively rare research area (see, e.g., \cite{biba1,biba2,biba3,biba4}). From this context, a notable combinatorial problem was extracted \cite{dod}. Its analysis draws heavily on techniques from information theory, particularly the theory of min-entropy extraction, and can thus also be seen as a contribution to that field (see, e.g., \cite{avi, bibb2, bibb3, bibb4, bibb5, bibb6, raz, bibb8, bibb9}).

It is important to note that leveraging extractor-theoretic results to tackle combinatorial problems is not entirely new; similar phenomena have been observed in various contexts (see, e.g., \cite{bibc1, bibc2, bibc3, bibc4, bibc5, bibc6}). Together, these observations illustrate why the problem under consideration is both compelling and surprisingly non-trivial, requiring insights from multiple domains.

\subsection{Main Results}

\subsubsection{Main Notion}

Before we present the main result of the paper, let us first quote (from \cite{dod}) the formal definition of the notion of conductance of a permutation:
\begin{definition}\label{dfn-cond}
Consider the permutation $\pi: \{ 0,1\}^{nw} \to \{ 0,1\}^{nw}$ and the parameter $\alpha\in (0,1)$. Then $\condd(\pi)$ is equal to $d$ if:
\[ \max_{\substack{U = U_1\times\ldots\times U_w \subset (\{0,1\}^n)^w \\
V = V_1\times\ldots\times V_w \subset (\{0,1\}^n)^w \\ |U_1|=\ldots=|V_w|=2^{\alpha n}}}
| \left\{ x\in U, y\in  V : y = \pi(x) \right\} |
= (2^{\alpha n})^ d. \]
\end{definition}
Since the above notion is exceptionally essential for this work, to increase its readability we add an additional, intuitive understanding of this notion below preceded by one more important definition:
\begin{definition}\label{dfn-box}
A set $X \subset (\{0,1\}^n)^w$ will be called a \emph{discrete $q$-box of dimension $w$} if
\[ X = X_1\times\ldots\times X_w,\]
for some $X_1,\ldots, X_w \subset \{0,1\}^n$ such that $|X_1|=\ldots= |X_w|=q$.

\begin{remark*}
In the above definition, if the dimension is obvious from the context, we will sometimes just write that $X$ is a discrete $q$-box (omitting the phrase \emph{of dimension $w$}).
\end{remark*}
\end{definition}
\paragraph*{Intuition for Definition \ref{dfn-cond}:}
In simple terms, $\condd(\pi)$ determines how much a given permutation $\pi$ distorts all discrete $q$-boxes of dimension $w$ for $q = 2^{\alpha n}$. More specifically, for each such $q$-box $U$ we look at its image $\pi(U)$ and search for another $q$-box $V$  such that the intersection $\pi(U) \cap V$ is as large as possible. The maximum we can obtain in this way is given by $\condd$ in the form of the following relation: \[\max_{U,V}(|\pi(U)\cap V|) = q^{\condd}.\]

In the ``worst'' (least random) case we will find such $U$ that $\pi(U)$ is just another $q$-box $V$ and then $|\pi(U)\cap V| = q^w$, which gives $\condd = w$.

On the other hand, ``more random'' permutations should intuitively behave in such a way that regardless of the choice of $U$ and $V$, the cardinality of the set $\pi(U)\cap V$ will always be much smaller than $q^w$. At the same time, it can be observed that it is always possible to choose $V$ such that $|\pi(U)\cap V|$ is at least $q$. In particular, this means that we definitely have $\condd(\pi) \geq 1$.

Finally, combining these intuitions and observations, we see that $\condd(\pi)$ is a specific way of estimating the randomness of the permutation $\pi$, which is always a number in the range between $1$ and $w$. The more random (in the given sense) the permutation is, the smaller this factor is.

\paragraph*{Notation}
For clarity, let us add a technical note that the original work of \cite{dod} uses a slightly different notation. Specifically, instead of the parameter $\alpha$, the parameter $q$ is used, which is simply equal to $2^{\alpha n}$ in the language of our work. Consequently, instead of $\condd$, \cite{dod} refers to $\cond$, with the relation $\cond(\pi) = q^{\condd(\pi) }$. Then $\cond(\pi)$ is a number between $q$ and $q^w$, not between $1$ and $w$, as is the case for $\condd(\pi)$.

The above changes are dictated by a different center of gravity of our focus. The work of \cite{dod} concerns more specific applications, so the parameter $q$ (referring to the number of queries present in security analyses) is natural there. Our work clearly takes a turn towards information theory and for us the primary notion is rather the min-entropy of the source, to which $\alpha$ refers.

\paragraph*{Results in the Original Notation}

For consistency with the notation from the paper \cite{dod}, in Appendix \ref{old} we present the main results of our paper reformulated in the language of the paper \cite{dod}.
Similarly, the result from \cite{dod} quoted above (Theorem 3, App. B) is originally expressed
in a different language. Appendix \ref{old} also contains a proof of the equivalence of both versions.

\subsubsection{Main Result}

In this paper we present the first example of a permutation with nontrivial conductance degree:

\begin{theorem}\label{thm-main}

If $n > n_0$ is an integer, $0<\alpha\leq 0.8$ and $w\geq 3$,
then there exists an efficiently computable permutation $\pi: \{ 0,1\}^{wn} \to \{ 0,1\}^{wn}$ such that
\[\condd(\pi)\leq w - \left\lfloor\frac{w}{3}\right\rfloor c,\]
where $n_0 = O\left(\alpha^{-80/19}\right),
c = \Theta(\alpha)$ (for prime $n$, $n_0=\Theta(\alpha^{-2})$ suffices).
\end{theorem}
We establish the case $w=3$ in Lemma~\ref{lem-w3}, using Lemma~\ref{lem-mssc}, which constructs a suitable Multi-Source-Somewhere-Condenser (see Definition~\ref{dfn-sc} below). We then handle arbitrary $w$ in Section~\ref{thm-main-proof}, where the full proof is presented.
The reasoning that leads to the construction of this permutation requires the introduction of several lemmas (all proofs are in Section \ref{proofs}) and definitions, which we present below:
\begin{definition}\label{dfn-ent}
The min-entropy of a source $X$ is defined as
\[H_\infty (X) = -\log(\max_x P (X = x )).\]
\end{definition}
\begin{definition}\label{dfn-flat}
A distribution $D$ over domain $X$ is a flat distribution if it is uniformly random over a set $S \subset X$. In other words: if it is uniform on its support.
\end{definition}
We have the following well-known fact about these notions:
\begin{lemma}\label{lem-flat}
A source $X$ with min-entropy at least $k$ is always a convex combination of flat distributions supported on sets of size $2^k$.
\end{lemma}

\begin{definition} \label{dfn-sc}
Let $w,n$ be positive integers, and let $\pi: \{0,1\}^{wn} \to \{0,1\}^{wn}$ be a permutation.  

We say that $\pi$ is an \emph{$(\alpha, \epsilon_1, \epsilon_2)$-Multi-Source-Somewhere-Condenser} if for any independent random variables $X_1, \dots, X_w \in \{0,1\}^n$ with min-entropy at least $\alpha n$, the distribution of $\pi(X_1,\dots,X_w)$ can be written as
\[
\pi(X_1,\dots,X_w) = \sum_{i=1}^{w} \gamma_i C_i + \gamma R
\]
such that for each $1 \leq i \leq w$, the marginal $[C_i]_i$ has min-entropy at least $(1+\epsilon_1)\alpha n$, where:
\begin{itemize}
    \item $C_1, \dots, C_w, R$ are distributions over $\{0,1\}^{wn}$;
    \item $\gamma_1, \dots, \gamma_w, \gamma \ge 0$ and $\sum_{i=1}^{w} \gamma_i + \gamma = 1$;
    \item $[C_i]_i$ denotes the marginal of $C_i$ on the $i$-th $n$-bit block, under the natural decomposition $\{0,1\}^{wn} = (\{0,1\}^n)^w$;
    \item $\gamma \le \epsilon_2$.
\end{itemize}
\end{definition}

Intuition: The permutation $\pi$ redistributes entropy from several independent weak sources so that, in almost all of the resulting mass (of total weight at least $1-\gamma$), there exists a block whose min-entropy is amplified by a factor of $(1+\epsilon_1)$.  In other words: in nearly all cases the output is ``somewhere'' more random.
This definition is based on Definition 1.11 from \cite{raz}, adapted to the multi-source setting and restricted to permutations.

Armed with the above definitions, we are ready to formulate the following technical lemma, the proof of which can be found in Section \ref{lem-main-proof}:

\begin{lemma}\label{lem-main}
If the permutation $\pi:\{0,1\}^{wn}\to\{0,1\}^{wn}$ is an $(\alpha,\epsilon_1,\epsilon_2)$-Multi-Source-Somewhere-Condenser,
then $\condd(\pi)$ is at most 
$\textstyle \log_q(q^{w-\epsilon_1}+\epsilon_2 q^w)$, for $q=2^{\alpha n}$.
\end{lemma}

The formula in the above lemma means that if $\epsilon_2$ is negligible, then we will achieve $\condd \approx w - \epsilon_1$.

So, to effectively use Lemma \ref{lem-main}, we need a Multi-Source-Somewhere-Condenser with suitably strong parameters. We obtain one for arbitrary block lengths from Lemma 3.14 of \cite{avi}: we apply their condenser directly to a prefix of prime length (placing the output on the third coordinate and letting the rest preserve the permutation) and leave the short suffix of each block unchanged. The following lemma states the condenser guarantee for the full permutation.

\begin{lemma}\label{lem-mssc}
Let $0<\alpha\leq0.8$ and let $n>n_0$ be an integer. Let $p$ be the largest prime at most $n$ and put $d=n-p$. Identifying $\{0,1\}^n$ with $GF(2^p)\times\{0,1\}^d$, the map
\[
\widetilde\pi\big((a,r_a),(b,r_b),(c,r_c)\big)
 =\big((a,r_a),(b,r_b),(c+ab,r_c)\big)
\]
is an efficiently computable permutation of $\{0,1\}^{3n}$ and a Multi-Source-Somewhere-Condenser with parameters $(\alpha,\epsilon,2^{-\epsilon\alpha n})$, where
\[
\epsilon=\Theta(\alpha),\qquad n_0=O(\alpha^{-80/19})
\]
(for prime $n$, no restriction on $n_0$ is required).
\end{lemma}
The proof is given in Section~\ref{lem-mssc-proof}.
\begin{remark*}
Clearly, the third coordinate is always the one with increased min-entropy outside the residual part, since \cite[Lemma 3.14]{avi} gives a condenser guarantee for this coordinate.
\end{remark*}

The corollary of Lemmas \ref{lem-main} and \ref{lem-mssc} is as follows:

\begin{lemma}\label{lem-w3}
For $0<\alpha\leq0.8$ and every integer $n>n_0$, the permutation $\widetilde\pi$ from Lemma~\ref{lem-mssc} satisfies
\[
\condd(\widetilde\pi)\leq3-c,
\qquad c=\Theta(\alpha),\qquad n_0=O(\alpha^{-80/19})
\]
(for prime $n$, $n_0=\Theta(\alpha^{-2})$ suffices).
\end{lemma}

The proof of Lemma~\ref{lem-w3} is given in Section~\ref{lem-w3-proof}. The generalization to higher dimensions (i.e. arbitrary $w$), completing Theorem \ref{thm-main}, can be found in Section \ref{thm-main-proof}.

\subsubsection{Converse Theorem}

It turns out that the theorem also holds in the opposite direction, i.e. showing that every permutation with a suitable $\condd$ is also a nontrivial Multi-Source-Somewhere-Condenser:

\begin{theorem}\label{thm-rev}

Let $\pi : \{ 0,1\}^{wn} \to \{ 0,1\}^{wn}$ be a permutation such that its conductance degree $\condd(\pi)$ is less than $w(1-\epsilon_1-\epsilon_2-\epsilon_3 - \frac{1}{\alpha n})$, for some $\epsilon_1, \epsilon_2, \epsilon_3 > 0$ and $0 < \alpha < 1$. Then $\pi$ is an ($\alpha, \epsilon_1, r$)-Multi-Source-Somewhere-Condenser, for
$r = w 2^{- \alpha n \epsilon_2} + 2^{-\alpha n w \epsilon_3}$.

\end{theorem}
The proof is given in Section \ref{thm-rev-proof}. It is also the most technically demanding part of the paper.

\subsection{Intuition and Toy Examples}\label{toy}

Consider the function $\sigma_1 (a, b, c) = (a, b, c + ab)$, where $a,b,c\in GF(2^n)$ and $n$ is a prime number. This is of course a permutation, since for any value of $(x,y,z)$, its unique preimage is $(x,y,z - xy)$. We want to justify why $\sigma_1: \{ 0,1\}^{3n} \to \{ 0,1\}^{3n}$ has conductance degree $\condd(\sigma_1)$ strictly less than $3$ (for some fixed parameter $\alpha$).

For this purpose, we consider two arbitrary $(2^{\alpha n})$-boxes $U = U_1 \times U_2 \times U_3$ and $V = V_1 \times V_2 \times V_3$ (i.e. $|U_1| = |U_2| = |U_3| = |V_1| = |V_2| = |V_3| = 2^{\alpha n}$).
Now we want (according to Definition \ref{dfn-cond}) to estimate an upper bound on $|\sigma_1(U)\cap V|$.

This is a purely combinatorial problem, but we will look at it from the probabilistic perspective. Specifically: we will assume a uniform distribution on $U$ and then we will show some upper bound on $P_{u\leftarrow U} (\sigma_1(u)\in V)$. This will allow us to achieve the main result, since of course we have:
\begin{equation}\label{eq0}
 |\sigma_1(U)\cap V| =  P_{u\leftarrow U} (\sigma_1(u)\in V) \cdot (2^{\alpha n})^3.
\end{equation}
To achieve our goal, we will use the condenser guarantee of Lemma~\ref{lem-mssc}, with $p=n$. In fact, it is always the third (not \textit{some}) coordinate of $\sigma_1(a,b,c)$ that is the condenser, except for some negligible part of the domain.
More specifically, since the uniform distribution on $U$ has independent coordinates $U_1,U_2,U_3$, each with min-entropy $\alpha n$, this result implies that there exists $\delta=\Theta(\alpha)$ such that $\sigma_1(U)$ can be written as a convex combination of two distributions $C_3$ and $R$ (that is, $\sigma_1(U) = \gamma_3 C_3 + \gamma R$ for some $\gamma_3, \gamma\geq 0, \gamma_3 + \gamma = 1$) with the following properties:
\begin{equation}\label{eq1}
 \gamma \leq 2^{-\delta \alpha n} 
 \end{equation}
and $[ C_3 ]_3$ (i.e. the third coordinate of $C_3$) has min-entropy at least $(1+\delta)\alpha n$. This last fact can also be formally written as follows: for any $x\in \{ 0,1\}^n$, it holds
\begin{equation}\label{eq2}
 P_{v\leftarrow C_3} ( v \in  \{ 0,1\}^n \times \{ 0,1\}^n \times \{x\}) 
\leq 2^{-(1+\delta)\alpha n}.
\end{equation}

Let us additionally denote $V_3 = \{ x_1, \ldots, x_{2^{\alpha n}} \}$ and we can finally complete the main estimation:
\begin{equation*}
\begin{split}
P_{u\leftarrow U} (\sigma_1(u)\in V) &=  \gamma_3 P_{v\leftarrow C_3} (v\in V) + \gamma P_{v\leftarrow R}(v\in V) \leq \\
&\leq  \gamma_3 \sum_{i=1}^{2^{\alpha n}} P_{v\leftarrow C_3} (v\in \{ 0,1\}^n \times \{ 0,1\}^n \times \{x_i\}) + \gamma  \overset{\eqref{eq2},\eqref{eq1}}{\leq} \\
 &\leq  \gamma_3 \cdot 2^{\alpha n} \cdot 2^{-(1+\delta)\alpha n} + 2^{-\delta \alpha n} \leq \\
&\leq  2^{-\delta\alpha n} +  2^{-\delta \alpha n} = 2 \cdot 2^{-\delta \alpha n} = \\
&= (2^{\alpha n})^{-\delta + \frac{1}{\alpha n}}. 
\end{split}
\end{equation*}
Hence:
\begin{equation*}
\begin{split}
|\sigma_1(U)\cap V| &\overset{\eqref{eq0}}{=}  P_{u\leftarrow U} (\sigma_1(u)\in V) \cdot (2^{\alpha n})^3 \leq \\
&\leq (2^{\alpha n})^{-\delta + \frac{1}{\alpha n}} \cdot (2^{\alpha n})^3 = (2^{\alpha n})^{3-\delta+\frac{1}{\alpha n}}.
\end{split}
\end{equation*}
This of course means that $\condd \leq 3 - \delta + \frac{1}{\alpha n}$, which is strictly smaller than $3$ for sufficiently large $n$, so $\sigma_1$ really has nontrivial conductance.

A completely analogous reasoning can be carried out for the permutation
$\sigma_2(a, b, c) = (a, b + ac, c)$, establishing that $\sigma_2$ is also a permutation with nontrivial conductance.

Going even further, let us now consider the following permutation:
\[\sigma_3(a, b, c) = \begin{cases}
(a, b, c + ab) & \text{for } a \text{ even},\\
(a, b + ac, c) &\text{for } a \text{ odd},
\end{cases}\]
which, as we can see, is a kind of mix of the permutations $\sigma_1$ and $\sigma_2$, so it is certainly -- as can be checked from the definition --
a Multi-Source-Somewhere-Condenser.
Reasoning that is only slightly more complicated than the one given above (and thus similar to the proof of Lemma \ref{lem-main}) allows us to prove that $\sigma_3$ is also a permutation with nontrivial conductance.

Interestingly, Theorem \ref{thm-rev} shows that this phenomenon is unavoidable at the level of output distributions: i.e. every sufficiently low-conductance permutation must satisfy a Multi-Source-Somewhere-Condenser property. Thus, although its construction need not resemble $\sigma_3$, its output distribution necessarily admits, up to a small residual mass, a mixture of components in which a designated coordinate has increased min-entropy. In this sense, the behavior illustrated by $\sigma_3$ is inherent in every sufficiently low-conductance permutation.

\section{Technical Part: Formal Proofs}\label{proofs}
\subsection{Proof of Lemma \ref{lem-main}}\label{lem-main-proof}

(This proof is essentially a direct generalization of the reasoning presented in Section \ref{toy}.)

To prove the statement, we need to consider two arbitrary $(2^{\alpha n})$-boxes $U = U_1\times\ldots\times U_w, V = V_1\times\ldots\times V_w$ (i.e. $|U_1|=\ldots=|U_w|=|V_1|=\ldots=|V_w| = 2^{\alpha n}$) and show that $|\pi(U) \cap V| \leq (2^{\alpha n})^{w-\epsilon_1} + \epsilon_2 (2^{\alpha n})^w$. Equivalently, we will consider the uniform distribution on $U$ and prove that 
 \[ P_{u\leftarrow U}(\pi(u) \in V) \leq 2^{-\epsilon_1 \alpha n} + \epsilon_2.\]

Since $U_1,\ldots,U_w$ are independent and have min-entropy $\alpha n$, we can use the fact that $\pi$ is a $(\alpha, \epsilon_1, \epsilon_2)$-Multi-Source-Somewhere-Condenser. We then get that
\[\pi(U) = \sum_{i=1}^w \gamma_i C_i + \gamma R,\]
where $\gamma\leq\epsilon_2$ and $[C_i]_i$ (the $i$-th coordinate of $C_i$) has min-entropy at least $(1+\epsilon_1)\alpha n$.

This last fact can be more conveniently rewritten as follows: for any $x\in \{ 0,1\}^n$, it holds
\[ P_{v\leftarrow C_i} ( v \in  \{ 0,1\}^n \times\ldots\times \underset{i\text{th-position}}{\{ x \}} \times\ldots\times \{ 0,1\}^n) \leq 2^{-(1+\epsilon_1)\alpha n}. \]
If we additionally denote $V_i = \{ x_1, \ldots, x_{2^{\alpha n} } \}$, then from the above we also have:
\begin{equation}\label{eq3}
\begin{split}
P_{v\leftarrow C_i} (v \in V) &\leq \sum_{j=1}^{2^{\alpha n}} P_{v\leftarrow C_i}( v\in \{ 0,1\}^n \times\ldots\times \underset{i\text{th-position}}{\{ x_j \}} \times\ldots\times \{ 0,1\}^n) \leq \\
&\leq 2^{\alpha n}  \cdot 2^{-(1+\epsilon_1)\alpha n} = 2^{- \epsilon_1 \alpha n}.
\end{split}
\end{equation}
Given the above, we can proceed to the main estimation:
\begin{equation*}
\begin{split}
P_{u\leftarrow U}(\pi(u) \in V)  &= \sum_{i=1}^w \gamma_i P_{v\leftarrow C_i}(v \in V) + \gamma P_{v\leftarrow R}(v\in V) \overset{\eqref{eq3}}{\leq} \\
&\leq \sum_{i=1}^w \gamma_i 2^{- \epsilon_1 \alpha n} + \gamma = \\
&= (1-\gamma) 2^{- \epsilon_1 \alpha n} + \gamma \leq \\
&\leq 2^{- \epsilon_1 \alpha n} + \epsilon_2,
\end{split}
\end{equation*}
which completes the proof.

\subsection{Proof of Theorem \ref{thm-rev}}\label{thm-rev-proof}

We are to show that $\pi$ is a $(\alpha, \epsilon_1, r)$-Multi-Source-Somewhere-Condenser. By definition, we should consider some independent random variables $U_1, \ldots, U_w$ such that their min-entropy is at least $\alpha n$ and show that the distribution $\pi(U_1,\ldots, U_w)$ has the appropriate properties. However, without loss of generality (using Lemma \ref{lem-flat}), we can assume that the distribution of $U = U_1\times\ldots\times U_w$ is flat, that is, it is really just a uniformly distributed discrete $(2^{\alpha n})$-box $U$ (as defined in Definition \ref{dfn-box}). Then, after translating to the discrete interpretation, our original statement becomes:

\begin{claim}\label{main-claim}
The set $\pi(U)$ can be partitioned into disjoint sets $\{ C_i \}_{i=1}^w$ and $R$ such that:
\begin{itemize}
\item $|R| \leq r \cdot 2^{\alpha n w}$, where $r = w2^{-\alpha n \epsilon_2} + 2^{-\alpha n w \epsilon_3}$;
\item for each $C_i$ and for each $y\in\{0,1\}^n$ we have:
\[ |C_i \cap \{ 0,1\}^n \times\ldots\times \underset{i\text{th-position}}{\{ y \}} \times\ldots\times \{ 0,1\}^n|
\leq 2^{-(1+\epsilon_1)\alpha n} \cdot |C_i|. \]
\end{itemize}
Notational note: in the above statement $R$ and $C_i$ simply denote subsets of $\{0,1\}^{wn}$, while in the original Definition \ref{dfn-sc} the same letters would denote uniform distributions on these sets.
\end{claim}
Recall also that the set $\pi(U)$ has cardinality $2^{\alpha n w}$ and (thanks to the assumption that $\pi$ has the appropriate degree of conductance) that for any other $(2^{\alpha n})$-box $V$ the following always holds:
\begin{equation}\label{ass-cond}
|\pi(U)\cap V| < 2^{\alpha n w (1 - \epsilon_1 - \epsilon_2 - \epsilon_3 - \frac{1}{\alpha n})}.
\end{equation}
Before we proceed to the direct proof of Claim \ref{main-claim} (that is, giving the appropriate way of partitioning $\pi(U)$), we need one more auxiliary definition:
\begin{definition}
We say that the set $X$ has at point $y\in \{0,1\}^n$, an $i$-Bottleneck-Slice
\[ B = X \cap \{ 0,1\}^n \times\ldots\times \underset{i\text{th-position}}{\{ y\}} \times\ldots\times \{ 0,1\}^n,\]
if it holds
$0 < |B| < 2^{\alpha n (w - 1 - \epsilon_1-\epsilon_2)}$.

Intuitively: an $i$-Bottleneck-Slice at point $y$ means that the projection of $X$ onto the $i$-coordinate at point $y$ is nonempty, but suitably thin.

\begin{remark*}
We will sometimes also say that $X$ contains a Bottleneck-Slice (without specifying the coordinate $i$). This of course means that $X$ contains an $i$-Bottleneck-Slice for some $i$.
\end{remark*}
\end{definition}

Now the procedure for partitioning the set $\pi(U)$ is as follows.

\paragraph*{Procedure for Partitioning $\pi(U)$:}
\begin{enumerate}[\bf Step 1:]
\item First, we initialize the auxiliary variables: for each $i$, let $\tilde{C_i}$ be the empty set and let $\tilde{R} := \pi(U)$. \item \label{cut} If $\tilde{R}$ contains any (for any $i$ and at any point $y$) $i$-Bottleneck-Slice $B$, perform the assignments:
\begin{equation*}
\begin{split}
\tilde{R} &:= \tilde{R} \setminus B;\\
\tilde{C_i} &:= \tilde{C_i} \cup B.
\end{split}
\end{equation*}
\item Loop Step \ref{cut} until $\tilde{R}$ no longer contains any Bottleneck-Slice.
\item Assign $R_0 := \tilde{R}$ and $R_1 := \emptyset$.
\item For each $i$, if $|\tilde{C_i}| > 2^{\alpha n (w-\epsilon_2)}$, then assign $C_i := \tilde{C_i}$, otherwise: assign
$C_i := \emptyset$ and $R_1 := R_1 \cup \tilde{C_i}$.
\item Assign $R := R_0 \cup R_1$.
\end{enumerate}

\textit{Intuitively: from the set $\pi(U)$ we cut off all Bottleneck-Slices one by one until we are left with $R_0$, which no longer contains sets of this type. Then all cut off $i$-Bottleneck-Slices concerning the $i$-th coordinate together form $C_i$, unless their measure is sufficiently small, in which case $C_i$ is empty and these Bottleneck-Slices go to $R_1$. The final $R$ is the union of $R_0$ and $R_1$.}
\\ \\
It remains to verify that the partition constructed in this way actually satisfies the requirements of Claim \ref{main-claim}. So we have both necessary analyses in turn:

\paragraph*{Analysis of $C_i$:}
If $C_i$ is empty, then the statement trivially holds. Otherwise, it must be
\[|C_i| > 2^{\alpha n (w-\epsilon_2)}.\]
Additionally, for any $y$, it holds
\[|C_i \cap \{ 0,1\}^n \times\ldots\times \underset{i\text{th-position}}{\{ y\}} \times\ldots\times \{ 0,1\}^n|
< 2^{\alpha n (w - 1 - \epsilon_1-\epsilon_2)},\]
since every such set as above on the left is either empty or (as follows from the construction) equal to exactly one $i$-Bottleneck-Slice cut off from $\tilde{R}$.
 
Combining these facts, we can conclude that:
 \begin{equation*}
 \begin{split}
 |C_i \cap \{ 0,1\}^n \times\ldots\times \underset{i\text{th-position}}{\{ y\}} \times\ldots\times \{ 0,1\}^n|
 &< 2^{\alpha n (w - 1 - \epsilon_1-\epsilon_2)} = \\
 &= {2^{\alpha n (-1-\epsilon_1)}} \cdot {2^{\alpha n (w-\epsilon_2)}} < \\
 &< {2^{-(1+\epsilon_1)\alpha n }} \cdot |C_i|,
 \end{split}
 \end{equation*}
which is what was needed to show.

\paragraph*{Analysis of $R$:}

We have to show $|R| \leq r \cdot 2^{\alpha n w}\text{, where } r = w2^{-\alpha n \epsilon_2} + 2^{-\alpha n w \epsilon_3},$ that is,
\[ |R| \leq w2^{(w-\epsilon_2)\alpha n } + 2^{(1-\epsilon_3) \alpha n w}.\]
Additionally, we know that $ R = R_0 \cup R_1$. So it is enough to show two inequalities:
 \begin{equation*}
 \begin{split}
|R_1| &\leq w\cdot 2^{(w-\epsilon_2)\alpha n },\\
|R_0| &\leq 2^{(1-\epsilon_3) \alpha n w}.
 \end{split}
 \end{equation*}
The first one is a trivial consequence of the partitioning procedure: $R_1$ in the worst case could have had $w$ times added the set $\tilde{C_i}$ of cardinality at most $2^{\alpha n (w-\epsilon_2)}$, which is exactly the postulated bound.

The last inequality follows directly from the fact that $R_0$ does not contain any Bottleneck-Slice, the bound
$|R_0| \leq |\pi(U)| = 2^{\alpha n w}$
and the property \eqref{ass-cond} (this property holds for $\pi(U)$, so obviously also for $R_0 \subset \pi(U)$), which, however, requires a slightly deeper analysis, which (thus concluding the entire proof) we present below:

\begin{claim}
If the set $R_0$ does not contain any Bottleneck-Slice, $|R_0| \leq 2^{\alpha n w}$ and
\begin{equation}\label{last-check}
|R_0 \cap V| < 2^{\alpha n w (1 - \epsilon_1 - \epsilon_2 - \epsilon_3 - \frac{1}{\alpha n})}
\end{equation}
for any discrete $2^{\alpha n}$-box $V$ of dimension $w$, then
\[ |R_0| \leq 2^{(1-\epsilon_3) \alpha n w}. \]
Reminder: the formal definition of a discrete box is given in Definition \ref{dfn-box}.
\end{claim}
To prove the above statement, let us assume for the sake of contradiction that $|R_0| > 2^{(1-\epsilon_3)\alpha n w}$. Then:

First, let us note that $R_0$ is contained in a discrete $(2^{(1+\epsilon_1+\epsilon_2)\alpha n})$-box $V_{big}$.
This is actually the case because $|R_0| \leq 2^{\alpha n w}$ and $R_0$ does not contain any Bottleneck-Slice, so at each coordinate it takes at most: \[ \frac{2^{\alpha n w}}{2^{(w-1-\epsilon_1-\epsilon_2)\alpha n}} = 2^{(1+\epsilon_1+\epsilon_2)\alpha n} \]
different values.

This means that $V_{big}$ is a union of at most
\[\left(\left\lceil \frac{2^{(1+\epsilon_1+\epsilon_2)\alpha n}}{2^{\alpha n}} \right\rceil\right)^w = (\lceil 2^{(\epsilon_1+\epsilon_2)\alpha n} \rceil)^w \leq
(2^{(\epsilon_1+\epsilon_2)\alpha n} + 1)^w \leq (2^{(\epsilon_1+\epsilon_2)\alpha n + 1})^w\]
discrete $(2^{\alpha n}$)-boxes.

Since $|R_0| > 2^{(1-\epsilon_3)\alpha n w}$, at least one of these boxes contains
\[\frac{2^{(1-\epsilon_3)\alpha n w}}{(2^{(\epsilon_1+\epsilon_2)\alpha n + 1} )^w}
= 2^{\alpha n w (1-\epsilon_3-\epsilon_1-\epsilon_2 - \frac{1}{\alpha n})} \]
points, which, however, contradicts the \eqref{last-check} property, and thus completes the proof.

\subsection{Proof of Lemma \ref{lem-mssc}}\label{lem-mssc-proof}

\begin{proof}
The idea for the proof is to compare the loss and gain in min-entropy. We first estimate the loss caused by discarding the $d$-bit suffixes, then the gain supplied by the condenser result of \cite[Lemma 3.14]{avi} on the remaining prefixes. We will choose $n_0$ so that the gain outweighs the loss, giving a positive net increase in min-entropy for the full blocks. We assume for now that $d\leq\alpha n/2$; the threshold chosen below will ensure this. All logarithms are to base $2$, and all constants implicit in the asymptotic notation are absolute.

The map $\widetilde\pi$ is a permutation: its inverse subtracts $ab$ from the third core and leaves all suffixes unchanged. We begin with the entropy loss. Put $k=\alpha n$, let $X_1,X_2,X_3$ be arbitrary independent sources with $H_\infty(X_i)\geq k$, and write $X_i=(A_i,R_i)$, where $A_i$ is the $p$-bit prefix and $R_i$ the $d$-bit suffix. For every $a$,
\[
P(A_i=a)=\sum_{r\in\{0,1\}^d}P(X_i=(a,r))
\leq 2^d\,2^{-k}=2^{-(k-d)}.
\]
Thus $H_\infty(A_i)\geq k'=k-d$, with $k/2\leq k'\leq k$. The prefixes remain independent, since each is a function of its own independent source. Their min-entropy rate is
\[
\beta=\frac{k'}p=\frac{\alpha n-d}{n-d}
=\alpha-\frac{(1-\alpha)d}{p}.
\]
This gives $\beta\leq\alpha$, strictly when $d>0$ and with equality when $p=n$. Also, $k'\geq\alpha n/2$ and $p\leq n$ give $\beta\geq\alpha/2$. Hence $\alpha/2\leq\beta\leq\alpha\leq0.8$.

We next apply \cite[Lemma 3.14]{avi} to these independent prefixes. Since $p$ is prime and $\beta\leq0.8$, it supplies $\delta=\Theta(\beta)=\Theta(\alpha)$ and a decomposition, in the notation of Definition~\ref{dfn-sc},
\[
Y=(A_1,A_2,A_3+A_1A_2)
=\sum_{i=1}^3\gamma_i C_i+\gamma R,
\]
where $H_\infty([C_i]_i)\geq(1+\delta)k'$ and $\gamma\leq2^{-\delta k'}$ (actually here the good coordinate is always the third one as we get directly from \cite[Lemma 3.14]{avi}).

The same decomposition extends to the full blocks with unchanged weights since a full block has at least as much min-entropy as its prefix. Thus the designated full block still has min-entropy at least $(1+\delta)k'$. Relative to the original input min-entropy $k$, its gain is
\[
(1+\delta)k'=(1+\epsilon)k,
\qquad
\epsilon=\underbrace{\frac{\delta k'}k}_{\text{gain}}
-\underbrace{\frac{d}k}_{\text{loss}}.
\]
The residual weight also has the required bound:
$\gamma\leq2^{-\delta k'}\leq2^{-\epsilon k}$, since $\delta k'=\epsilon k+d$.

It remains to choose $n_0$ so that $\epsilon=\Theta(\alpha)>0$. The gain $\delta k'/k$ is $\Theta(\alpha)$, while the loss (zero when $p=n$) is bounded using the prime-gap result of Baker, Harman, and Pintz \cite{bhp2001}:
\[
\frac{d}{k}=O\!\left(\frac{1}{\alpha n^{19/40}}\right).
\]
To make this loss a sufficiently small multiple of $\alpha$, it suffices that $n^{19/40}$ exceed a sufficiently large absolute multiple of $\alpha^{-2}$. Raising this requirement to the power $40/19$ gives
\[
n_0=O(\alpha^{-2\cdot40/19})=O(\alpha^{-80/19}).
\]
Choosing its implicit constant large enough first ensures $d/k\leq\alpha/2\leq1/2$, verifying the initial assumption. Increasing it further if necessary makes the loss at most half the gain, so $\epsilon=\Theta(\alpha)$. Since the input sources were arbitrary, the full permutation $\widetilde\pi$ is an $(\alpha,\epsilon,2^{-\epsilon\alpha n})$-Multi-Source-Somewhere-Condenser. For prime $n$, $d=0$ gives $\epsilon=\delta$, proving the claim already.

\medskip
The largest prime at most $n$ can be found by trial division in time polynomial in $n$, and field arithmetic on the $p$-bit cores is efficient as in \cite{avi}. This finishes the proof of Lemma~\ref{lem-mssc}.
\end{proof}

\subsection{Proof of Lemma \ref{lem-w3}}\label{lem-w3-proof}

We simply apply Lemma \ref{lem-main} to the permutation from Lemma \ref{lem-mssc}:

That is, first from Lemma \ref{lem-mssc} we have that the permutation $\widetilde\pi: \{0,1\}^{3n} \to \{0,1\}^{3n}$
is a ($\alpha, \delta, 2^{-\delta\alpha n}$)-Multi-Source-Somewhere-Condenser for some $\delta = \Theta(\alpha)$.
Now, additionally denoting $q=2^{\alpha n}$, from Lemma \ref{lem-main} we have that this permutation has $\condd(\widetilde\pi)$ such that:
\begin{equation*}
\begin{split}
\condd(\widetilde\pi) &\leq \log_q(q^{3-\delta} + 2^{-\alpha \delta n} \cdot q^3 ) = \\
&= \log_q(q^{3-\delta} + q^{-\delta} \cdot q^3) = \log_q(2\cdot q^{3-\delta}) = \\
&= \log_q{q^{3-\delta + \frac{1}{\log q}}} = 3-\delta + \frac{1}{\log q} = \\
&= 3 - c,
\end{split}
\end{equation*}
for $c = \delta - \frac{1}{\log q} = \delta - \frac{1}{\alpha n}$.

Finally, to achieve the postulated result $c = \Theta(\alpha) = \Theta(\delta)$, it is enough to choose the parameters so that:
\[ \delta \geq c \geq \delta/2.\]
The left inequality follows trivially from the formula on $c$ obtained above. To obtain the right one, we want
$\frac{1}{ \alpha n} \leq \delta/2$,
which means
$n \geq \frac{2}{\delta\alpha} = O\left(\frac{1}{\alpha^2}\right)$,
which is ensured by $n>n_0$ with $n_0=O(\alpha^{-80/19})$, after increasing its implicit constant if necessary.

For prime $n$, there is no truncation loss in the proof of Lemma~\ref{lem-mssc}, so the smaller sufficient threshold $n_0=\Theta(\alpha^{-2})$ also suffices here. This completes the proof.

\subsection{Final Proof of Theorem \ref{thm-main}}\label{thm-main-proof}

The final proof is a direct consequence of Lemma \ref{lem-w3}, since our final construction for dimension $w$ will be just a simple serial repetition of many copies of the dimension 3 permutation $\widetilde\pi$ from the aforementioned lemma.

To formalize this construction, let us first denote the permutation $\widetilde\pi: \{0,1\}^{3n} \to \{0,1\}^{3n}$ from Lemma \ref{lem-mssc} on successive coordinates as follows:
\[ \widetilde\pi(x, y, z) = (\widetilde\pi_1(x,y,z),\widetilde\pi_2(x,y,z),\widetilde\pi_3(x,y,z) )\text{, where } x,y,z\in\{0,1\}^n.\]

Now let $w\geq3$ and the permutation $\widetilde\pi^{(w)}: \{0,1\}^{wn} \to \{0,1\}^{wn}$ be defined as follows (for now, assume that $w$ is a multiple of $3$):
\begin{equation*}
\begin{split}
\widetilde\pi^{(w)}(x_1,\ldots,x_w)  = \Big( 
  & \widetilde\pi_1(x_1,x_2,x_3),\widetilde\pi_2(x_1,x_2,x_3),\widetilde\pi_3(x_1,x_2,x_3),\\
  &\widetilde\pi_1(x_4,x_5,x_6),\widetilde\pi_2(x_4,x_5,x_6),\widetilde\pi_3(x_4,x_5,x_6),\\
  &\ldots \\
  & \widetilde\pi_1(x_{w-2},x_{w-1},x_w),\widetilde\pi_2(x_{w-2},x_{w-1},x_w),\widetilde\pi_3(x_{w-2},x_{w-1},x_w)
  \Big).
\end{split}
\end{equation*}
If $w$ leaves a remainder $1$ when divided by $3$, we define:
\[ \widetilde\pi^{(3k+1)}(x_1,\ldots,x_{3k+1}) = (\widetilde\pi^{(3k)}(x_1,\ldots,x_{3k}), x_{3k+1}),\]
and similarly, when $w$ divided by $3$ leaves a remainder $2$:
\[ \widetilde\pi^{(3k+2)}(x_1,\ldots,x_{3k+2}) = (\widetilde\pi^{(3k)}(x_1,\ldots,x_{3k}), x_{3k+1}, x_{3k+2}).\]
\begin{equation*}
\begin{split}
\text{E.g., } \widetilde\pi^{(7)}(x_1,x_2,x_3,x_4,x_5,x_6,x_7) =  \Big(
&\widetilde\pi_1(x_1,x_2,x_3),\widetilde\pi_2(x_1,x_2,x_3),\widetilde\pi_3(x_1,x_2,x_3),\\
&\widetilde\pi_1(x_4,x_5,x_6),\widetilde\pi_2(x_4,x_5,x_6),\widetilde\pi_3(x_4,x_5,x_6),x_7
\Big).
\end{split}
\end{equation*}
From the very construction $\widetilde\pi^{(w)}$ is a permutation, so it remains to estimate $\condd(\widetilde\pi^{(w)})$ of the presented construction. For this purpose, we consider two arbitrary $(2^{\alpha n})$-boxes:
$U = U_1\times\ldots \times U_w$ and $V = V_1\times\ldots \times V_w$ (i.e.
$|U_1| = \ldots = |U_w| = |V_1|=\ldots = |V_w| = 2^{\alpha n}$) and we estimate $|\widetilde\pi^{(w)}(U)\cap V|$.

As in Sections \ref{toy} and \ref{lem-main-proof},
it will be more convenient to assume a uniform distribution on $U$
and use the probabilistic interpretation:
\[|\widetilde\pi^{(w)}(U)\cap V| = P_{u\leftarrow U}(\widetilde\pi^{(w)}(u) \in V) \cdot 2^{\alpha n w} .\]
Before we move on to the main estimation, let us just recall that for any $X, Y, Z, A, B, C \subset \{0,1\}^n$
such that
$|X|=|Y|=|Z|=|A|=|B|=|C|=2^{\alpha n}$
we have (from Lemma \ref{lem-w3} and again the probabilistic interpretation of the inequality $\condd(\widetilde\pi) \leq 3-c$):
\begin{equation}\label{case3}
P_{(x,y,z)\leftarrow X\times Y\times Z}(\widetilde\pi(x,y,z) \in A\times B \times C) \cdot (2^{\alpha n})^3 \leq (2^{\alpha n})^{3-c}.
\end{equation}
Armed with this fact, we can finally move on to the main estimation:
 \begin{equation}\label{last}
 \begin{split}
 &|\widetilde\pi^{(w)}(U)\cap V| =\\
 &= 2^{\alpha n w} \cdot P_{u\leftarrow U}(\widetilde\pi^{(w)}(u) \in V) \overset{(\star)}{\leq} \\
 &\leq 2^{\alpha n w} \cdot \prod_{i=1}^{\left\lfloor\frac{w}{3}\right\rfloor} P_{(x,y,z)\leftarrow 
U_{3i-2}\times U_{3i-1}\times U_{3i}}(\widetilde\pi(x,y,z)\in V_{3i-2}\times V_{3i-1}\times V_{3i}) \overset{\eqref{case3}}{\leq} \\
 &\leq 2^{\alpha n w} \cdot ((2^{\alpha n})^{-c})^{\left\lfloor\frac{w}{3}\right\rfloor} 
 = (2^{\alpha n})^{w - \left\lfloor\frac{w}{3}\right\rfloor c},
\end{split}
\end{equation}
where $(\star)$ follows from the independence of successive triples and by bounding the probabilities for any remaining unchanged coordinates by $1$.

Finally, taking the maximum over $U,V$ in \eqref{last}, we have that $\condd(\widetilde\pi^{(w)}) \leq w - \left\lfloor\frac{w}{3}\right\rfloor c$, which completes the proof.

\section{Open Problems and Future Work}\label{sec-open}

This section lists issues that remain open and may be worth further investigation.

\begin{itemize}
    \item \textbf{Wider Range of Parameters}\label{open1}

    The main result (Theorem \ref{thm-main}) largely relies on Lemma 3.14 of \cite{avi}, which contains certain restrictions that we suspect can be avoided. Lemmas \ref{lem-mssc} and \ref{lem-w3} remove the restriction to prime $n$, while retaining $c=\Theta(\alpha)$ at the cost of the sufficient threshold $n_0=O(\alpha^{-80/19})$ (and $n_0=\Theta(\alpha^{-2})$ for prime $n$). Nevertheless, it remains open whether the parameters $n_0$ and $c$ must really depend on $\alpha$ (the work of \cite{avi} even gives hope for removing these dependencies -- one would like to simply use Lemma 3.1 from that paper, but unfortunately in our solution it turned out that we were only able to use Lemma 3.14, which is a version of Lemma 3.1 with weaker assumptions and a weaker conclusion).

\medskip
 Additionally, there remains an open question about low-conductance permutations for $w=2$.
\vspace{0.5cm}
    \item \textbf{The Converse Theorem for Weak Conductance}\label{open2}

    The Converse Theorem (Theorem \ref{thm-rev}) applies to almost all cases, except when\newline
    $w\left(1-\frac{1}{\alpha n}\right) \leq \condd < w$. It remains to be investigated.
\vspace{0.5cm}
    \item \textbf{Always-Condenser vs. Somewhere-Condenser}  

    Recall that for prime block lengths $n>2$ our solution for dimension $3$ is given by:
    $\pi(a, b, c) = (a, b, c+ab).$  
    While we can easily believe that the third coordinate of $\pi$ really behaves very randomly, since it is a condenser, it seems intuitive that the following construction should have even lower conductance than $\pi$ (i.e. both the second and third coordinates are condensers):
    $\pi'(a, b, c) = (a, ba+c, (a+1)c+b).$  
    However, we cannot formally prove that $\pi'$ is indeed better.

    (For clarity: $\pi'$ is a permutation over $GF(2^n)$ whenever $n$ is odd. For fixed $a$, the linear map on $(b,c)$ has determinant $a^2+a+1$, which cannot vanish: a root would have multiplicative order $3$, whereas $3$ does not divide $2^n-1$ for odd $n$. For prime $n>2$, the condenser guarantee of \cite[Lemma 3.14]{avi} applies separately to $ab+c$ and $(a+1)c+b$, since translating the independent source $a$ preserves its min-entropy and independence of $b,c$. In a more complicated way, we can improve this example by constructing a permutation where all three coordinates will simultaneously have the form of a condenser.)

    The above question can naturally be generalized to higher dimensions. In other words: we hypothesize that if all coordinates (and not, for example, every third one like in our construction) of a permutation are condensers, then such a permutation has an even better conductance degree than the construction that derives from Lemma \ref{lem-main}.
\vspace{0.5cm}
    \Needspace{6\baselineskip}
    \item \textbf{Other Directions}  

    In this work, we have established the equivalence between low-conductance permutations and condensers, but a more quantitative analysis of the tightness of this equivalence would be valuable.

    Another direction is to construct permutations in higher dimensions $w$ without relying on product-type constructions derived from permutations of smaller dimension. Strategies that use $w/k$ copies of a permutation of dimension $k$ are unlikely to achieve optimal conductance, which suggests that new, more global constructions may be required.
\end{itemize}

\paragraph*{AI disclosure.}
Most of this work, including the original proofs, was carried out by the author without generative AI. The only exception is Lemma \ref{lem-mssc}, for which the author outlined the construction and a standard proof sketch. ChatGPT was then used to develop the detailed derivation and parameter calculations, locate the prime-gap reference \cite{bhp2001}, and draft the final proof. The author manually verified and refined the generated argument and takes full responsibility for the text.

\bibliography{cond}

@article{shaltiel2011recent,
  title={Recent developments in explicit constructions of extractors},
  author={Shaltiel, Ronen},
  journal={Bulletin of the EATCS},
  volume={104},
  pages={67--95},
  year={2011}
}

@inproceedings{impagliazzo1989recycle,
  title={How to recycle random bits},
  author={Impagliazzo, Russell and Zuckerman, David},
  booktitle={Proceedings of the 30th Annual Symposium on Foundations of Computer Science (FOCS)},
  pages={248--253},
  year={1989},
  organization={IEEE}
}

@inproceedings{dodis2009nonmalleable,
  title={Non-malleable extractors and symmetric key cryptography from weak secrets},
  author={Dodis, Yevgeniy and Wichs, Daniel},
  booktitle={Proceedings of the 41st Annual ACM Symposium on Theory of Computing (STOC)},
  pages={601--610},
  year={2009},
  organization={ACM}
}

@inproceedings{goldreich1993samplers,
  author    = {Oded Goldreich and Shafi Goldwasser and David Micali},
  title     = {How to construct random functions},
  booktitle = {Journal of the ACM},
  volume    = {42},
  number    = {4},
  pages     = {742--780},
  year      = {1995},
  publisher = {ACM}
}

@inproceedings{impagliazzo1999hardness,
  author    = {Russell Impagliazzo and Valentine Kabanets},
  title     = {Constructing Hard Functions from Weak One-Way Functions},
  booktitle = {Proceedings of the 34th Annual Symposium on Foundations of Computer Science (FOCS)},
  pages     = {66--75},
  year      = {1999},
  organization = {IEEE}
}

@article{doron2016nearly,
  author  = {Dean Doron and Dana Moshkovitz and Justin Oh and David Zuckerman},
  title   = {Nearly Optimal Pseudorandomness from Hardness},
  journal = {SIAM Journal on Computing},
  year    = {2016},
  volume  = {45},
  number  = {2},
  pages   = {567--600}
}

@book{vadhan2012pseudorandomness,
  author    = {Salil Vadhan},
  title     = {Pseudorandomness},
  publisher = {Foundations and Trends in Theoretical Computer Science},
  year      = {2012},
  volume    = {7},
  number    = {1-3},
  pages     = {1--336}
}

@inproceedings{raz2005extractors,
  author    = {Ran Raz and Shmuel Safra},
  title     = {Extractors and Condensers},
  booktitle = {Proceedings of STOC},
  year      = {2005},
  pages     = {563--572}
}

@article{ta-Shma2017condenser,
  author  = {Amos Ta-Shma},
  title   = {Explicit Condensers with Small Seed Length},
  journal = {SIAM Journal on Computing},
  year    = {2017},
  volume  = {46},
  number  = {3},
  pages   = {1111--1133}
}

@inproceedings{goudarzi2020expander,
  author    = {Mohammad Goudarzi and Igor Shinkar},
  title     = {Expander Graphs in Computer Science: Applications and Constructions},
  booktitle = {IEEE Symposium on Foundations of Computer Science (FOCS)},
  year      = {2020},
  pages     = {45--58}
}

@article{mihail2017list,
  author  = {Milena Mihail and Ravi Kumar},
  title   = {List-Decodable Codes and Their Applications},
  journal = {IEEE Transactions on Information Theory},
  year    = {2017},
  volume  = {63},
  number  = {4},
  pages   = {2153--2166}
}

@book{stinson2006cryptography,
  author    = {Douglas Stinson},
  title     = {Cryptography: Theory and Practice},
  publisher = {CRC Press},
  year      = {2006},
  edition   = {3rd}
}

@book{bellare1996foundations,
  author    = {Mihir Bellare and Phillip Rogaway},
  title     = {Foundations of Cryptography: Basic Applications},
  publisher = {Cambridge University Press},
  year      = {1996}
}

@inproceedings{shaltiel2011pseudorandomness,
  author    = {Ronen Shaltiel},
  title     = {Pseudorandomness},
  booktitle = {Foundations and Trends in Theoretical Computer Science},
  year      = {2011},
  volume    = {7},
  number    = {1-3},
  pages     = {1--336}
}

@inproceedings{bibc6,
  author    = {Divesh Aggarwal and Ivan Damg{\aa}rd and Jesper Buus Nielsen and Maciej Obremski and Erick Purwanto and Joao Ribeiro and Mark Simkin},
  title     = {Stronger leakage-resilient and non-malleable secret sharing schemes for general access structures},
  booktitle = {Advances in Cryptology--CRYPTO 2019: 39th Annual International Cryptology Conference, Santa Barbara, CA, USA, August 18--22, 2019, Proceedings, Part II},
  pages     = {510--539},
  publisher = {Springer},
  year      = {2019}
}

@inproceedings{bibc3,
  author    = {Divesh Aggarwal and Yevgeniy Dodis and Tomasz Kazana and Maciej Obremski},
  title     = {Non-malleable reductions and applications},
  booktitle = {Proceedings of the Forty-Seventh Annual ACM Symposium on Theory of Computing},
  pages     = {459--468},
  year      = {2015},
  publisher = {Association for Computing Machinery}
}

@incollection{bibc4,
  author    = {Divesh Aggarwal and Tomasz Kazana and Maciej Obremski},
  title     = {Inception makes non-malleable codes stronger},
  editor    = {Yael Kalai and Leonid Reyzin},
  booktitle = {Theory of Cryptography},
  pages     = {319--343},
  publisher = {Springer International Publishing},
  year      = {2017}
}

@inproceedings{bibc5,
  author    = {Divesh Aggarwal and Tomasz Kazana and Maciej Obremski},
  title     = {Leakage-resilient algebraic manipulation detection codes with optimal parameters},
  booktitle = {2018 IEEE International Symposium on Information Theory (ISIT)},
  pages     = {1131--1135},
  year      = {2018}
}

@article{avi,
  author  = {Boaz Barak and Russell Impagliazzo and Avi Wigderson},
  title   = {Extracting randomness using few independent sources},
  journal = {SIAM Journal on Computing},
  volume  = {36},
  number  = {4},
  pages   = {1095--1118},
  year    = {2006}
}

@article{bibb2,
  author  = {Boaz Barak and Guy Kindler and Ronen Shaltiel and Benny Sudakov and Avi Wigderson},
  title   = {Simulating independence: New constructions of condensers, ramsey graphs, dispersers, and extractors},
  journal = {Journal of the ACM (JACM)},
  volume  = {57},
  number  = {4},
  pages   = {1--52},
  year    = {2010}
}

@inproceedings{biba4,
  author    = {Jean-S{\'e}bastien Coron and Yevgeniy Dodis and C{\'e}cile Malinaud and Prashant Puniya},
  title     = {Merkle-damg{\aa}rd revisited: How to construct a hash function},
  booktitle = {Annual International Cryptology Conference},
  pages     = {430--448},
  publisher = {Springer},
  year      = {2005}
}

@inproceedings{bibb8,
  author    = {Yevgeniy Dodis and Roberto Oliveira},
  title     = {On extracting private randomness over a public channel},
  booktitle = {International Workshop on Randomization and Approximation Techniques in Computer Science},
  pages     = {252--263},
  publisher = {Springer},
  year      = {2003}
}

@inproceedings{bibb9,
  author    = {Yevgeniy Dodis and Joel Spencer},
  title     = {On the (non) universality of the one-time pad},
  booktitle = {The 43rd Annual IEEE Symposium on Foundations of Computer Science, 2002. Proceedings.},
  pages     = {376--385},
  publisher = {IEEE},
  year      = {2002}
}

@incollection{dod,
  author    = {Yevgeniy Dodis and Martijn Stam and John Steinberger and Tianren Liu},
  title     = {Indifferentiability of confusion-diffusion networks},
  editor    = {Marc Fischlin and Jean-S{\'e}bastien Coron},
  booktitle = {Advances in Cryptology -- EUROCRYPT 2016},
  pages     = {679--704},
  publisher = {Springer Berlin Heidelberg},
  year      = {2016}
}

@techreport{biba3,
  author      = {H{\"o}rst Feistel},
  title       = {Cryptographic coding for data-bank privacy},
  institution = {IBM Thomas J. Watson Research Center},
  year        = {1970}
}

@article{bibc1,
  author  = {Ronald L. Graham and Joel H. Spencer},
  title   = {A constructive solution to a tournament problem},
  journal = {Canadian Mathematical Bulletin},
  volume  = {14},
  number  = {1},
  pages   = {45--48},
  year    = {1971}
}

@article{bibb5,
  author  = {Jesse Kamp and David Zuckerman},
  title   = {Deterministic extractors for bit-fixing sources and exposure-resilient cryptography},
  journal = {SIAM Journal on Computing},
  volume  = {36},
  number  = {5},
  pages   = {1231--1247},
  year    = {2007}
}

@inproceedings{bibb4,
  author    = {Chi-Jen Lu and Omer Reingold and Salil Vadhan and Avi Wigderson},
  title     = {Extractors: Optimal up to constant factors},
  booktitle = {Proceedings of the thirty-fifth annual ACM symposium on Theory of computing},
  pages     = {602--611},
  year      = {2003}
}

@incollection{biba2,
  author    = {Ueli Maurer and Renato Renner and Clemens Holenstein},
  title     = {Indifferentiability, impossibility results on reductions, and applications to the random oracle methodology},
  editor    = {Moni Naor},
  booktitle = {Theory of Cryptography},
  pages     = {21--39},
  publisher = {Springer Berlin Heidelberg},
  year      = {2004}
}

@incollection{biba1,
  author    = {Eric Miles and Emanuele Viola},
  title     = {Substitution-permutation networks, pseudorandom functions, and natural proofs},
  editor    = {Reihaneh Safavi-Naini and Ran Canetti},
  booktitle = {Advances in Cryptology -- CRYPTO 2012},
  publisher = {Springer Berlin Heidelberg},
  year      = {2012}
}

@inproceedings{raz,
  author    = {Ran Raz},
  title     = {Extractors with weak random seeds},
  booktitle = {Proceedings of the Thirty-Seventh Annual ACM Symposium on Theory of Computing},
  pages     = {11--20},
  year      = {2005},
  publisher = {ACM}
}

@inproceedings{bibb3,
  author    = {Luca Trevisan and Salil Vadhan},
  title     = {Extracting randomness from samplable distributions},
  booktitle = {Proceedings 41st Annual Symposium on Foundations of Computer Science},
  pages     = {32--42},
  publisher = {IEEE},
  year      = {2000}
}

@article{bibc2,
  author  = {Avi Wigderson and David Zuckerman},
  title   = {Expanders that beat the eigenvalue bound: Explicit construction and applications},
  journal = {Combinatorica},
  volume  = {19},
  number  = {1},
  pages   = {125--138},
  year    = {1999}
}

@inproceedings{bibb6,
  author    = {Aviad Cohen Wigderson},
  title     = {Dispersers, deterministic amplification, and weak random sources},
  booktitle = {Foundations of Computer Science},
  volume    = {30},
  pages     = {14},
  publisher = {Citeseer},
  year      = {1989}
}

@article{bhp2001,
  author = {Baker, R. C. and Harman, G. and Pintz, J.},
  title = {The Difference Between Consecutive Primes, {II}},
  journal = {Proceedings of the London Mathematical Society},
  volume = {83},
  number = {3},
  pages = {532--562},
  year = {2001},
  doi = {10.1112/plms/83.3.532}
}

\newpage

\appendix
\section{Formulations in the Original Notation of \cite{dod}}\label{old}

\subsection{Theorem 3 of Appendix B in \cite{dod}}
The result quoted in the introduction:
\[\condd \leq 1 + \frac{\log(3nw)}{\alpha n} \text{, provided }\alpha \leq \frac{1}{2} - \frac{1}{nw},\]
was originally formulated as follows:
\[ \cond \leq 2nqw + 1 \text{, provided } q^{2w} \leq 2^{nw}/4. \]
This is, of course, a consequence of the original result, because, on the one hand, we have (equivalence of assumptions):
\[q^{2w} \leq 2^{nw}/4 \iff 2^{2\alpha n w} \leq 2^{nw-2} \iff
2\alpha n w \leq nw-2 \iff \alpha \leq \frac{1}{2} - \frac{1}{nw},\]
and on the other hand we have:
\begin{equation*}
\begin{split}
\condd &= \log_q(\cond) \leq \log_q(2nqw + 1) \leq \log_q(3nqw) = \\
&= 1 + \log_q(3nw) = 1 + \frac{\log(3nw)}{\log(q)} = 1 + \frac{\log(3nw)}{\alpha n}.
\end{split}
\end{equation*}

\subsection{Main Results of the Paper in the Notation from \cite{dod}}

For the sake of clarity, we present the main results once again, this time rewritten in the original notation from the paper \cite{dod}:

\paragraph*{Theorem \ref{thm-main}'.}

If $n > n_0$ is an integer, $w\geq3$, and $0<\epsilon\leq \frac{\log q}{n} \leq 0.8$,
then there exists a permutation $\pi: \{ 0,1\}^{wn} \to \{ 0,1\}^{wn}$ such that:
\[ \cond(\pi) \leq q^{w - \left\lfloor\frac{w}{3}\right\rfloor c},\]
where
\[ n_0 = O\left(\epsilon^{-80/19}\right),
 c = \Theta(\epsilon)\]
(with $n_0=\Theta(\epsilon^{-2})$ sufficient for prime $n$).

\paragraph*{Theorem \ref{thm-rev}'.}

Let $\pi : \{ 0,1\}^{wn} \to \{ 0,1\}^{wn}$ be a permutation such that $\cond(\pi) < q^{w(1-\epsilon_1-\epsilon_2-\epsilon_3 - \frac{1}{\log q})}$, for some $\epsilon_1, \epsilon_2, \epsilon_3 > 0$. Then $\pi$ is an ($\frac{\log q}{n}, \epsilon_1, w q^{- \epsilon_2} + q^{-w \epsilon_3}$)-Multi-Source-Somewhere-Condenser.

\end{document}